\documentclass[aps,twocolumn,superscriptaddress]{revtex4}
\usepackage{graphicx}

\newcommand{\ket}[1] {\left| #1 \right\rangle}

\begin{document}

\title{Simulating systems of itinerant spin-carrying particles using arrays of superconducting qubits and
resonators}

\author{S. Ashhab}
\affiliation{Qatar Environment and Energy Research Institute,
Doha, Qatar}
\affiliation{Center for Emergent Matter Science, RIKEN, Wako-shi,
Saitama, Japan}

\date{\today}

% July 22, 2014

\begin{abstract}
We propose possible approaches for the quantum simulation of
itinerant spin-carrying particles in a superconducting
qubit-resonator array. The standard Jaynes-Cummings-Hubbard setup
considered in several recent studies can readily be
used as a quantum simulator for a number of relevant phenomena,
including the interaction with external magnetic fields and
spin-orbit coupling. A more complex setup where multiple qubits and
multiple resonator modes are utilized in the simulation gives a
higher level of complexity, including the simulation of particles
with high spin values and allowing more direct control on processes
related to spin-orbit coupling. This proposal could be implemented in
state-of-the-art superconducting circuits in the near future.
\end{abstract}

%PACS numbers:
%42.50.Pq: Cavity quantum electrodynamics; micromasers
%71.36.+c: Polaritons (including photon-phonon and photon-magnon interactions)
%03.67.Lx: Quantum computation architectures and implementations
%03.67.Ac: Quantum algorithms, protocols, and simulations
%85.25.-j: Superconducting devices

\maketitle

\section{Introduction}
\label{Sec:Introduction}

Two classic paradigms in physics, namely the Jaynes-Cummings (JC)
model describing the interaction between light and matter
\cite{QuantumOpticsBooks} and the Hubbard model describing
particle motion in a periodic structure
\cite{CondensedMatterBooks}, were recently combined to
produce the Jaynes-Cummings-Hubbard (JCH) model
\cite{HartmannGreentreeAngelakis,DengReview,SuperconductingJCHReviews,CarusottoReview,Georgescu,Paraoanu}.
In the JCH model, one deals with a regular array of harmonic
oscillators (e.g.~optical cavities), each containing a two-level
[or more generally few-level] quantum system (e.g.~an atom). The
particles whose properties and dynamics are studied in such a
situation are generally hybridized photonic/atomic excitations
that hop between the lattice sites in a Hubbard-model-like
picture. Numerous theoretical proposals have highlighted the
novelty and potential applications of the model 
\cite{HartmannSpinModels,ChoFQHE,Carusotto,KochSFMI,Makin,Leib,
KochTRSYmmetryBreaking,Tsomokos,Nunnenkamp,Zheng,Yang,Xiang,Kim,
Kulaitis,Kapit}, and initial experiments along the way to
demonstrating large-scale JCH arrays have been reported recently
\cite{Underwood,Raftery,McKay}.

The most commonly studied setup for the JCH model comprises an
array of cavities between which photons can hop and each of which
contains a two-level atom. The Hamiltonian describing the setup is
given by
\begin{eqnarray}
\hat{H} & = & \sum_i \left( \frac{\hbar\omega_a}{2} \hat{\sigma}_z^{(i)} +
\hbar\omega_0 \hat{a}_i^{\dagger}\hat{a}_i + g
\hat{\sigma}_x^{(i)} (\hat{a}_i+\hat{a}_i^{\dagger}) \right)
\nonumber \\ & & + J \sum_{\langle i,j \rangle} \left(
\hat{a}_i^{\dagger}\hat{a}_j + h.c. \right),
\label{Eq:JCH_Hamiltoian}
\end{eqnarray}
where $\omega_a$ is the atomic resonance frequency, $\omega_0$ is
the cavity resonance frequency, $g$ is the atom-cavity coupling
strength, and $J$ is the inter-cavity coupling strength (or in
other words the inter-cavity hopping strength). The operators
$\hat{\sigma}_{\alpha}^{(i)}$ (with $\alpha=x$, $y$ or $z$) are
the Pauli matrices operating on the state of the two-level atom,
while $\hat{a}_i$ and $\hat{a}_i^{\dagger}$ are, respectively, the
annihilation and creation operators of the cavity field; the
indices $i$ and $j$ denote the location of the atom or cavity in
the lattice, and the notation $\langle i,j \rangle$ indicates
nearest-neighbour hopping. In writing
Eq.~(\ref{Eq:JCH_Hamiltoian}), we have assumed that the different
parameters are uniform across the entire lattice, and we have
assumed that there is no direct coupling between the atoms
\cite{Huemmer}.

One interesting property of the JCH model is the fact that in
typical setups the number of excitations is not conserved; in most
proposals excitations are to be created by coherently driving the
cavities, thus creating quantum superpositions of states with
different total particle numbers. As a result, one typically needs
to consider the balance between the injection of excitations into
the system by the externally applied fields (which can be pulsed
or continuous) and the loss of excitations through various
radiative or dissipative processes \cite{Carusotto}. As such, one
deals not with an equilibrium system, but rather a driven
dissipative system. This situation differs drastically from
commonly studied condensed-matter-physics problems (where particle
number is assumed to be fixed throughout an experiment, and losses
are typically treated as a limitation of a given experimental
setup). This difference could be seen as a disadvantage of the
proposed JCH setups when viewed as alternative quantum simulators,
e.g.~to replace atomic gases in optical lattices
\cite{Georgescu,Bloch}, for simulating electrons in solids and
related systems. On the other hand, the difference between the JCH
model and other, well-established paradigms with conserved particle
number (e.g.~atoms in optical lattices) presents an opportunity
where the new paradigm could offer novel applications, and this
possibility is an active research area \cite{Carusotto}.
Furthermore, there is no fundamental difficulty in creating states
with a well-defined particle number in the JCH model;
Rabi-oscillation dynamics in the atoms in principle allows
the controllable creation of any desired total particle number, and
the suppression of radiative losses could allow for motional
thermalization of the fluid on a timescale that is short compared
to the lifetime of the particles.

There have been proposals for implementing the JCH model in a
variety of physical systems, including optical-frequency cavities
with implanted or electromagnetically trapped atoms
\cite{HartmannGreentreeAngelakis}, superconducting resonators and
qubits fabricated on a chip
\cite{KochSFMI,Leib,KochTRSYmmetryBreaking,Tsomokos,Nunnenkamp,Yang},
nanomechanical resonators \cite{Jacobs} and phonon cavities with
impurities in silicon \cite{Soykal}. The superconducting platform
is particularly promising from an experimental point of view.
Superconducting qubits, which serve as artificial atoms in the
model, and superconducting resonators are making rapid advances in
terms of their quantum coherence and the level of control and
addressability that they allow \cite{QubitReviews}. Coherent
coupling between qubits and resonators has been achieved in
various settings and has led to demonstrations of numerous
phenomena predicted from the theory of quantum optics
\cite{CircuitQEDReviews}. Furthermore, large arrays of
superconducting resonators can readily be fabricated on an
electronic chip with present-day technology
\cite{SuperconductingJCHReviews,Underwood,Raftery,McKay}. An
implementation of the JCH model in such circuits can realistically
be expected in the coming few years. We shall therefore focus on this
implementation of the model.

Studies on the JCH model have mainly focused on the case where
only one excitation mode per site play an active role in the
representation of the particles in the lattice. The result
is that one ends up dealing with a system of effectively spinless
particles. Even in this case, several interesting phenomena can be
obtained. Examples include the fractional quantum Hall effect
\cite{ChoFQHE}, unusual propagation behaviour \cite{Makin}, novel
disorder effects \cite{Kulaitis}, a rich phase diagram in the case
of ultrastrong atom-cavity coupling \cite{Zheng}, topological models
\cite{Xiang} and Dirac points \cite{Kim}.

The addition of a spin degree of freedom to the particles can be
expected to result in a variety of new phenomena that are absent
in the spinless case. In this context it is worth mentioning the
large number of phenomena encountered in the study of
Bose-Einstein condensates of spin-carrying particles
\cite{SpinorBEC}. In particular, novel quantum states have been
predicted in relation to the motion of spin-carrying atoms in
optical lattices \cite{Demler} and spin-orbit coupling in atomic
gases \cite{Stanescu,Cai,Zhao,Zhai}. In this paper we present possible
routes towards the addition of a spin degree of freedom to the effective
particles in the JCH model.

There have been proposals to simulate spins in a JCH-like setup in
the literature \cite{HartmannSpinModels}. However, the simulated
systems described stationary spins, as opposed to itinerant
spin-carrying particles. The possible use of polariton spin,
photon polarization or different modes in optical cavities in
order to introduce additional degrees of freedom has also been
discussed in the literature
\cite{DengReview,CarusottoReview,Rubo}, and Ref.~\cite{HartmannTwoComponentFluid} proposed the simulation of a
two-component JCH model using a four-level atom in each cavity.
However, in these proposals the number of particles in the different
components was conserved, and the addition of a spin degree of freedom
was therefore not truly realized. A related setup for the simulation
of Luttinger liquids and spin-charge separation corresponding to
hard-core repulsion in a continuous medium was proposed in
Ref.~\cite{AngelakisSpinChargeSeparation}. It should be noted
that these proposals generally rely on atoms with specific
energy-level structures, making them rather difficult to
implement in a superconducting architecture.

Here we analyze the possibility of constructing a superconducting
quantum simulator for itinerant particles that can hop between
neighbouring sites in a regular lattice and that possess an internal
(spin or pseudo-spin) degree of freedom. As will be discussed
in detail in Sec.~\ref{Sec:PossibleRoutes}, different approaches have
different levels of difficulty but also allow different levels of
complexity as quantum simulators.

\section{The desired Hamiltonian}
\label{Sec:DesiredHamiltonian}

The quintessential Bose-Hubbard Hamiltonian for spinless particles contains two terms: hopping and on-site interaction;
\begin{eqnarray}
\hat{H}_{\rm BH} & = & - t \sum_{\langle i,j \rangle}
\hat{b}_{i}^{\dagger} \hat{b}_{j} + U \sum_{i}
\hat{b}_{i}^{\dagger} \hat{b}_{i}^{\dagger} \hat{b}_{i}
\hat{b}_{i},
\label{Eq:BH_Hamiltonian_WithoutSpin}
\end{eqnarray}
where $t$ is the hopping strength, $U$ is the on-site interaction strength, and $\hat{b}$ and $\hat{b}^{\dagger}$ are, respectively, the annihilation and creation operators of the bosonic particles in the lattice. Constructing this Hamiltonian should be relatively straightforward in a superconducting architecture. The single-site energy spectrum of the JCH model, namely the JC ladder, exhibits a series of energy levels pairs. These pairs naturally define two excitation modes, the so-called lower- and upper-branch polariton modes. In the case of large detuning between the qubits and the resonators (i.e.~the so-called dispersive regime), the energy levels in either one of the two modes have a linear-plus-quadratic dependence on the number of excitation quanta \cite{DispersiveRegimeFootnote}. In the absence of inhomogeneous trapping potentials, the linear part does not have any physical consequences and can be ignored, leaving only the quadratic term, which plays the role of the on-site interaction term in Eq.~\ref{Eq:BH_Hamiltonian_WithoutSpin}. If we also take the hopping strength to be small compared to the qubit-resonator detuning, excitations can hop between neighbouring sites but they cannot jump from one polariton branch to the other. As a result, if we consider a system where only one of the two branches is populated, the Hamiltonian in Eq.~(\ref{Eq:JCH_Hamiltoian}) reduces to an effective Hamiltonian given by Eq.~(\ref{Eq:BH_Hamiltonian_WithoutSpin}).

Our aim is to design a JCH setup where the resulting effective
particles possess a spin degree of freedom. We would therefore like
to engineer an effective Bose-Hubbard Hamiltonian with an additional,
internal degree of freedom. A Hamiltonian that contains only hopping
and on-site interaction terms, i.e.~the generalization of
Eq.~(\ref{Eq:BH_Hamiltonian_WithoutSpin}), would read
\begin{eqnarray}
\hat{H}_{\rm BH,s} & = & - \sum_{\langle i,j \rangle,\alpha,\beta}
t_{\alpha,\beta} \hat{b}_{i,\alpha}^{\dagger} \hat{b}_{j,\beta}
\nonumber \\ & & + \sum_{i,\alpha,\beta,\gamma,\delta}
U_{\alpha,\beta,\gamma,\delta} \hat{b}_{i,\alpha}^{\dagger}
\hat{b}_{i,\beta}^{\dagger} \hat{b}_{i,\gamma} \hat{b}_{i,\delta},
\label{Eq:BH_Hamiltonian_WithSpin}
\end{eqnarray}
where $t_{\alpha,\beta}$ is a matrix of hopping strengths that describe
the process in which a particle with spin value $\beta$ hops to a
neighbouring site and spin value $\alpha$, and
$U_{\alpha,\beta,\gamma,\delta}$ is a tensor that describes the
strengths of various local interaction processes that transform a
pair of particles with spin values $\gamma$ and $\delta$ to spin
values $\alpha$ and $\beta$. As before, we have assumed that the
different parameters are uniform across the entire lattice. The
inclusion of nonuniform parameters, such as trapping potentials,
would be quite straightforward in a superconducting architecture.

A Hamiltonian describing a system of spin-carrying particles would
in general also contain a Zeeman term, which in the Hubbard model takes
the form
\begin{equation}
\hat{H}_{\rm Zeeman} = - \sum_{i,\alpha,\beta} g {\bf S}_{\alpha,\beta}
\cdot {\bf B} \; \hat{b}_{i,\alpha}^{\dagger} \hat{b}_{i,\beta},
\end{equation}
where $g$ is the Land\'e $g$ factor, ${\bf S}_{\alpha,\beta}$ is
the (vector) spin operator for the spin value being simulated, and
${\bf B}$ is the magnetic field. It should be noted that the
hopping term in Eq.~(\ref{Eq:BH_Hamiltonian_WithSpin})
contains spin-changing hopping terms that are relevant to
spin-orbit coupling, as we shall discuss below.

The interaction term in Eq.~(\ref{Eq:BH_Hamiltonian_WithSpin})
contains a tensor with several parameters, which means that
designing a system where all these parameters are tunable can be a
challenging task, and we shall not attempt to do that here. We focus
on the basic Hamiltonian that contains hopping and magnetic terms:
\begin{equation}
\hat{H} = - \sum_{\langle i,j \rangle,\alpha,\beta} t_{\alpha,\beta}
\hat{b}_{i,\alpha}^{\dagger} \hat{b}_{j,\beta} -
\sum_{i,\alpha,\beta} g {\bf S}_{\alpha,\beta} \cdot {\bf B}
\hat{b}_{i,\alpha}^{\dagger} \hat{b}_{i,\beta}.
\label{Eq:BH_Hamiltonian_WithSpin_NoInteractions}
\end{equation}
This Hamiltonian would already exhibit several features that do
not exist in the spinless case, as evidenced by the rich phase diagrams
discussed in Refs.~\cite{Stanescu}. As we shall see below, a system with
a spin-independent hard-core repulsion between the particles (i.e.~a
system where double occupancy of a single site is prohibited) occurs
naturally in one of the simple limits in the JCH setup. We shall therefore
have in mind a Hamiltonian with such an interaction term $\hat{H}_{\rm int}$
added:
\begin{equation}
\hat{H}_{\rm int} = U \sum_{i,\alpha,\beta} \hat{b}_{i,\alpha}^{\dagger}
\hat{b}_{i,\beta}^{\dagger} \hat{b}_{i,\beta} \hat{b}_{i,\alpha},
\end{equation}
where $U$ is a large coefficient that represents the strong on-site repulsion.

\section{Possible routes for adding spin degrees of freedom to the JCH setup}
\label{Sec:PossibleRoutes}

In this section we discuss two possible approaches that can be
used to introduce internal degrees of freedom to the particles in
the JCH setup. The first approach, namely the use of multiple polariton
branches, is rather straightforward but could nevertheless be used to achieve
demonstrations of various relevant phenomena. The second approach,
namely the inclusion of multiple qubits coupled to each resonator, would
allow more complex simulations. Two further alternative
(but less promising) approaches are presented in Appendix A.

\subsection{Polariton branches}
\label{SubSec:PolaritonBranches}

We first consider the JCH setup described in Sec.~\ref{Sec:Introduction} and whose Hamiltonian is given by Eq.~(\ref{Eq:JCH_Hamiltoian}). Although most studies consider the spinless case and implicitly
focus on only one excitation mode (or branch), this setup
inherently supports two excitation modes, namely the lower- and
upper-polariton modes. These modes could be used to represent the
$\ket{\uparrow}$ and $\ket{\downarrow}$ states of spin-1/2
particles \cite{HalfIntegerBosonFootnote}. One can therefore achieve
a quantum simulation of itinerant spin-1/2 (or more accurately
pseudospin-1/2) particles using the basic JCH setup.

We now consider how different terms in the Hamiltonian can be simulated, and we start with the hopping term. In this work the hopping is assumed to take place through the resonators. When the qubit-resonator detuning is large (i.e.~in the dispersive regime), one of the two polariton modes has an almost purely photonic character while the other has a large probability of the qubit being in its excited state. The effective hopping strength for the mostly photonic mode will then be much larger than that for the other mode, which is generally undesirable because it can create a natural separation in behaviour between the particles in the different spin states. Perhaps the simplest way to make the two hopping strengths comparable, or even equal, is to move away from the dispersive regime and consider the case of near or exact resonance between the qubits and resonators. In the case of exact resonance, the two hopping strengths are exactly equal.

\begin{figure}[h]
\includegraphics[width=4.0cm]{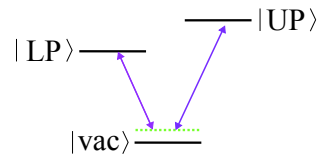}
\caption{Schematic diagram of a Raman process for
converting excitations between the two polariton modes in the JC
model. This process is used to simulate the Zeeman term in the
Hamiltonian, where the two states of a spin 1/2 particle mix
locally, i.e.~at a single site in the JCH lattice. The states
$\ket{\rm LP}$ and $\ket{\rm UP}$ are states with a single
excitation in the lower- and upper-polariton branches,
respectively. The state $\ket{\rm vac}$ is the vacuum state with
no excitations. The solid lines represent the three relevant energy
levels, and the dashed line represents a virtual energy level whose
location is determined by the detuning between the driving ac field
and the real energy levels.}
\label{Fig:ModeConversionRamanProcess}
\end{figure}

In order to fully realize the spin degree of freedom, we need to have processes that change the spin of a particle, e.g.~a Zeeman term in the Hamiltonian. In order to simulate the Zeeman term with magnetic fields pointing in arbitrary directions, we need to have the ability to induce the conversion of excitations between the lower- and upper-polariton modes. Since the two modes have different energies, one might think of achieving the conversion by driving the system at the frequency separation between the two modes. However, the relevant matrix elements for the driving fields that can be applied easily [with driving operators $(\hat{a}+\hat{a}^{\dagger})$ or $\hat{\sigma}_x$] vanish (because these operators change the excitation number while the excitation number is conserved in the Zeeman term), precluding the conversion between modes using one of these candidates for the driving field. The conversion could be achieved, however, using a Raman process with simultaneous driving at two frequencies, as illustrated in Fig.~\ref{Fig:ModeConversionRamanProcess}. Depending on the detuning between the driving field frequency difference and the actual energy level separation, one can obtain an effective magnetic field that points in any desired direction \cite{DrivingFieldPhaseFootnote}. In the case where the driving fields are chosen such that the Raman process occurs with perfect energy matching (while still having the detuning with the intermediate energy level), the effective magnetic field lies in the xy plane. Note that this (artificial) magnetic field couples only to the spin degree of freedom and does not lead to phenomena related to the coupling between the orbital motion and external magnetic fields, e.g.~the quantum Hall effect.

As was explained in detail in Ref.~\cite{Goldman}, spin-orbit coupling can be engineered by temporally alternating a lattice system between a few different Hamiltonian settings. These settings include only hopping terms and position dependent Zeeman terms, which are readily achievable in the superconducting architecture. For example, one can take the four-step sequence
\begin{widetext}
\begin{equation}
\left\{ \;\; \frac{\hat{p}_x^2}{m} \;\; , \;\; \frac{\hat{p}_x^2+\hat{p}_y^2}{2m}+\kappa(\hat{x}\hat{\sigma}_x-\hat{y}\hat{\sigma}_y) \;\; , \;\; \frac{\hat{p}_y^2}{m} \;\; , \;\; \frac{\hat{p}_x^2+\hat{p}_y^2}{2m}-\kappa(\hat{x}\hat{\sigma}_x-\hat{y}\hat{\sigma}_y) \;\; \right\},
\end{equation}
\end{widetext}
where $\hat{p}_x$ and $\hat{p}_y$ are the x and y components of the momentum operator in two dimensions and the operators $\hat{\sigma}_{\alpha}$ (with $\alpha=x,y$ or $z$) are the Pauli operators for the effective particles (Note that the quadratic dependence of energy on momentum can be obtained straightforwardly from the Hubbard model at the bottom of the energy band). If one applies this sequence repeatedly with a sufficiently high frequency $\omega$, one obtains the effective Hamiltonian
\begin{equation}
\hat{H}_{\rm eff} = \frac{\hat{p}_x^2+\hat{p}_y^2}{2m} + \lambda_{\rm R} (\hat{p}_x\hat{\sigma}_x+\hat{p}_y\hat{\sigma}_y) + \Omega_{\rm SO} \hat{L}_z\hat{\sigma}_z,
\label{Eq:StroboscopicEffectiveSOCHamiltonian}
\end{equation}
where $\lambda_{\rm R}=\pi\kappa/8m\omega$, $\Omega_{\rm SO}=-(8m/3)\lambda_{\rm R}^2$ and $\hat{L}_z=\hat{x}\hat{p}_y-\hat{y}\hat{p}_x$. This effective Hamiltonian is related to the standard Rashba form [i.e.~Eq.~(\ref{Eq:StroboscopicEffectiveSOCHamiltonian}) without the last term] by a unitary transformation \cite{Goldman}.

Theoretical studies predict a rich phase diagram for the mean-field ground state of a Bose gas with spin-orbit coupling, including stripe and vortex patterns \cite{Stanescu}. These states can be prepared using local operations on the different lattice sites, making them relatively easy to prepare in the proposed JCH setup with superconducting circuits. If one prepares a mean-field state that is close to the true ground state of the system (for a given particle number), the subsequent dynamics would exhibit only small amplitude changes. On the other hand, if the prepared initial state is far from the true ground state for a given set of parameters, one would expect that the state of the system will change drastically as it evolves in time. This way, one can test the theoretical predictions and possibly map out the ground-state phase diagram.

There are a few points that must be noted in this context. One of
the main issues that require care is the following:
if only the $\ket{\uparrow}$ state or the $\ket{\downarrow}$ state
is occupied at any given lattice site (i.e.~if only one of the two
polariton branches is populated at a given site), the energy-level
ladder allows multiple occupation of that site. A difficulty arises,
however, when dealing with quantum states that contain particles
in both states $\ket{\uparrow}$ and $\ket{\downarrow}$ at the same
site. The mapping between polariton excitations and spin-carrying
particles breaks down in this case, as can be seen by direct inspection of
the energy-level structure. For example, there are three different
spin states corresponding to a doubly occupied lattice site, whereas
there are only two states in the JC ladder in the two-excitation
subspace. The mapping is therefore only applicable when
dealing with systems where it suffices to consider at most a single
particle per site, which generally corresponds to dilute systems. Another difficulty is encountered if we take the hopping strength $J$ to be comparable to or larger than the single-site energy separation between the two polariton branches. In this case, the inter-cavity coupling dominates over the qubit-cavity coupling, and the separation between the two polariton modes becomes less clearly defined. In order to avoid this difficulty, the inter-cavity coupling must be kept weak compared to the polariton energy separation. Note that in this regime it becomes natural to say that particles in different spin states are strictly prohibited from occupying the same site while same-spin particles can do so at an effective energy cost (the latter having been analyzed in the literature on the JCH setup \cite{HartmannGreentreeAngelakis}). In order to simplify the problem and avoid dealing with these spin-dependent interactions, one can prohibit double occupancy altogether by choosing system parameters that correspond to the Mott-insulator regime. One then has hard-core on-site repulsion. It should be noted that even in this parameter regime the Mott insulator state is only realized for commensurate filling of the lattice. Away from these filling values, one has a system of itinerant particles.

To conclude this subsection, we summarize its main points. Using the basic JCH setup one can simulate a system of spin-1/2 particles where the tunneling strength can be made spin independent, an effective magnetic field can be engineered freely by driving the transition between the two polariton modes, and spin-orbit coupling can be obtained by periodically switching the Hamiltonian between four different settings. In the following subsection we consider a different approach that allows additional possibilities, in particular higher spin values and spin-orbit coupling with fixed rather than pulsed driving settings.

\subsection{Multiple qubits coupled to each resonator}
\label{SubSec:MultipleQubits}

\begin{figure}[h]
\includegraphics[width=9.0cm]{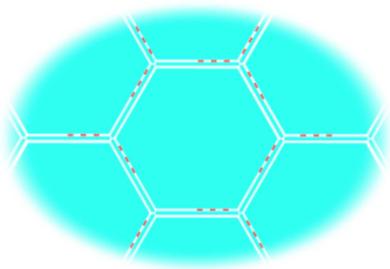}
\caption{Schematic diagram of a superconducting
resonator lattice with multiple qubits coupled to each resonator.
The cyan areas represent superconducting material, which is used
to define the resonators, while the red dots represent the qubits.
In this figure, the resonators are arranged in a Kagome lattice,
similarly to the setup studied in Refs.~\cite{KochTRSYmmetryBreaking}.
In this particular example, we include three qubits for each resonator,
which would be the case when simulating spin-1 particles. Note that the
placement of the qubits shown in this figure does not represent the
optimal placement; when designing an actual device, qubit placement is
typically determined based on the locations where the electric field in
the transmission line resonator has a large value, ideally a maximum.}
\label{Fig:JCHMultiQubitSchematic}
\end{figure}

We now consider the incorporation of multiple qubits
coupled to each resonator in a JCH system, as illustrated in Fig.~\ref{Fig:JCHMultiQubitSchematic}. Each qubit would then represent one
spin state of the simulated particles, keeping in mind that the
effective particles would actually be qubit-resonator hybridized
excitations. It is worth emphasizing here that this setup is not
unrealistic since there have been experiments with large arrays of
resonators as well as experiments with multiple qubits coupled to
a single resonator.

If we consider $N$ qubits coupled to one resonator mode, we find that there are in fact $N+1$ excitation modes. These modes can in principle by used to simulate a spin of size $s=N/2$. By using identical settings at all the lattice sites, one would obtain excitation modes with excitation frequencies that are distinct from each other but are uniform across the lattice. The different spin states of the simulated particles can then be identified through these frequencies. In this case, hopping processes arise naturally, and for sufficiently weak inter-cavity coupling, hopping processes do not change the spin state. However, for $N>2$ the occupation probability of the resonator cannot be equal for all the different singly excited states, which means that the hopping strengths for the different spin states cannot be made equal. In order to avoid this difficulty, one needs to employ multiple resonator modes. We therefore assume that we have $N$ qubits coupled to $N$ resonator modes at each site. Anticipating that we will use the resonator modes as mediators of various interactions, we generally assume that the parameters of each qubit-mode pair are in the dispersive regime and that the simulated particles are encoded in the qubit excitations.

Transitions between the different spin states, which are needed in
order to simulate magnetic fields and the Zeeman term in the
Hamiltonian, can be obtained by driving the qubits using any one of
the recently realized microwave-driving-based techniques for
implementing SWAP operations between superconducting qubits
\cite{Rigetti}. The resonators naturally provide an effective
inter-qubit coupling that can be used in these conversion protocols.
As discussed in Sec.~\ref{SubSec:PolaritonBranches}, the simulated
magnetic field can be designed to point in any direction depending on
the amplitude, phase and detuning of the driving fields.

\begin{figure}[h]
\includegraphics[width=8.0cm]{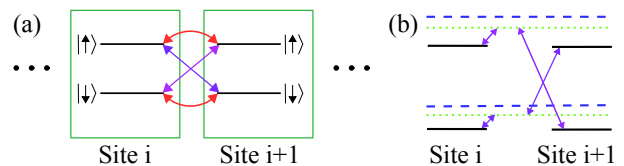}
\caption{Schematic diagrams illustrating the
processes involved in a spin-orbit coupled Hubbard model with spin-1/2 particles and how
these processes could be induced in a JCH system. The red arrows
describe spin-conserving hopping processes, while the purple
arrows describe spin-changing hopping processes. In (b) the black
solid lines describe the single-excitation qubit energy levels,
the blue dashed lines describe (delocalized) single-photon
energy levels in the resonators, and the green dotted lines show
virtual energy levels whose locations are determined by the detunings
between the ac fields and the real energy levels. The virtual energy
levels are somewhat detuned from the single-photon energy levels in
order to avoided populating the resonators with real excitations.}
\label{Fig:SpinOrbitCoupling}
\end{figure}

In order to simulate spin-orbit coupling, one needs to induce (controlled) spin-changing hopping processes. The
processes needed in the case of spin-1/2 particles are illustrated
in Fig.~\ref{Fig:SpinOrbitCoupling}a. These processes can be
realized by driving Raman transitions, as shown in
Fig.~\ref{Fig:SpinOrbitCoupling}b. One point that requires some
care here is that when the qubits are biased at their symmetry
points, convenient operators [such as
$(\hat{a}+\hat{a}^{\dagger})$ and $\hat{\sigma}_x$] have zero
matrix elements for the desired transitions. Modulating the qubit
frequencies [i.e.~driving the qubits using the operator
$\hat{\sigma}_z$], however, would induce these transitions
\cite{Porras}. If one wishes to achieve the maximum possible
coupling with this approach, one needs to work in the Landau-Zener
regime, where the qubit frequencies are driven up and down past the
resonator frequencies \cite{Shevchenko}. Depending on the system parameters, this requirement could imply strong modulation of the qubit frequency.
Another possibility for driving the Raman transitions without necessarily requiring strong driving on the qubits is using
tunable couplers between the qubits and the resonators and
modulating the coupling strength at the required frequencies
\cite{Niskanen,AshhabTunableCouplers}. The effective matrix
elements for the spin-changing hopping processes obtained this way
can be made on the order of the hopping strength $J$ (which occurs when
the driven qubit-resonator transitions have strengths comparable
to $J$), meaning that with the appropriate driving strengths
sufficiently large values of the spin-changing hopping matrix elements can be obtained. In order
to obtain a controllable form of spin-orbit coupling, one needs to
have the ability to control the amplitude and phase of the matrix
elements describing the spin-changing hopping processes. For
example, in order to obtain Rashba spin-orbit coupling in a square
lattice, the two spin-changing hopping matrix elements along one
of the two spatial dimensions need to have the same amplitude but
opposite signs \cite{Cai}. Similarly, spin-orbit coupling in one dimension can be obtained by designing the two spin-changing processes to
have opposite signs \cite{Zhao}. This sign difference can be
achieved by adjusting the phases of the ac fields that drive the
Raman transitions. This goal can be achieved with the proper choice
of parameters, as we show in the following derivation.

We now present a quantitative calculation showing how the different processes can arise in an effective Hamiltonian with the proper choice of parameters and driving conditions. For the simulation of spin-1/2 particles, where we only need to consider two resonator modes, the Hamiltonian of the entire system can be expressed as \cite{HatsAndHbarFootnote}
\begin{widetext}
\begin{eqnarray}
H & = & \sum_i \left( \omega_{\downarrow} c_{\downarrow,i}^{\dagger} c_{\downarrow,i} + \omega_{\uparrow} c_{\uparrow,i}^{\dagger} c_{\uparrow,i} + \omega_{r,1} a_{1,i}^{\dagger}a_{1,i} + \omega_{r,2} a_{2,i}^{\dagger}a_{2,i} + \sum_{s,m} g_{s,m,i}(t) \left( c_{s,i}^{\dagger} a_{m,i} + c_{s,i} a_{m,i}^{\dagger} \right) \right)
\nonumber \\ & & + \sum_m J_m \sum_{\langle i,j \rangle} \left( a_{m,i}^{\dagger} a_{m,i} + a_{m,i} a_{m,i}^{\dagger} \right).
\end{eqnarray}
Here we have introduced the annihilation and creation operators for qubit excitations $c$ and $c^{\dagger}$ with subscripts $\uparrow$ and $\downarrow$ as qubit labels. As above we use $a$ and $a^{\dagger}$ as the resonator annihilation and creation operators with subscripts $1$ and $2$ as mode labels. The different parameters in the Hamiltonian are self explanatory.

For this calculation we consider only two neighbouring lattice sites, and we focus on the single-excitation subspace. We also assume that the coupling strengths are modulated with sinusoidal time dependence, i.e.~$g(t)=g+f\cos(\omega t +\phi)$. Ignoring terms that do not have any significant effect on the system (e.g.~the dc component of the coupling between highly detuned subsystems), the Hamiltonian can be expressed as
\begin{eqnarray}
H & = & \omega_{\downarrow} \left( c_{\downarrow,i}^{\dagger} c_{\downarrow,i} + c_{\downarrow,j}^{\dagger} c_{\downarrow,j} \right) + \omega_{\uparrow} \left( c_{\uparrow,i}^{\dagger} c_{\uparrow,i} + c_{\uparrow,j}^{\dagger} c_{\uparrow,j} \right) + \omega_{r,1} \left( a_{1,i}^{\dagger} a_{1,i} + a_{1,j}^{\dagger} a_{1,j} \right) + \omega_{r,2} \left( a_{2,i}^{\dagger} a_{2,i} + a_{2,j}^{\dagger} a_{2,j} \right) \nonumber \\ & & + g_{\downarrow,1,i} \left( c_{\downarrow,i}^{\dagger} a_{1,i} + c_{\downarrow,i} a_{1,i}^{\dagger} \right) + g_{\downarrow,1,j} \left( c_{\downarrow,j}^{\dagger} a_{1,j} + c_{\downarrow,j} a_{1,j}^{\dagger} \right) + g_{\uparrow,2,i} \left( c_{\uparrow,i}^{\dagger} a_{2,i} + c_{\uparrow,i} a_{2,i}^{\dagger} \right) + g_{\uparrow,2,j} \left( c_{\uparrow,j}^{\dagger} a_{2,j} + c_{\uparrow,j} a_{2,j}^{\dagger} \right) \nonumber \\ & & + f_{\downarrow,1,i} \cos ( \omega_1 t + \phi_1 ) \left( c_{\downarrow,i}^{\dagger} a_{1,i} + c_{\downarrow,i} a_{1,i}^{\dagger} \right) + f_{\uparrow,1,j} \cos ( \omega_2 t + \phi_2 ) \left( c_{\uparrow,j}^{\dagger} a_{1,j} + c_{\uparrow,j} a_{1,j}^{\dagger} \right) \nonumber \\ & & + f_{\uparrow,2,i} \cos ( \omega_3 t + \phi_3 ) \left( c_{\uparrow,i}^{\dagger} a_{2,i} + c_{\uparrow,i} a_{2,i}^{\dagger} \right) + f_{\downarrow,2,j} \cos ( \omega_4 t + \phi_4 ) \left( c_{\downarrow,j}^{\dagger} a_{2,j} + c_{\downarrow,j} a_{2,j}^{\dagger} \right) \nonumber \\ & & + J_1 \left( a_{1,i}^{\dagger} a_{1,j} + a_{1,i} a_{1,j}^{\dagger} \right) + J_2 \left( a_{2,i}^{\dagger} a_{2,j} + a_{2,i} a_{2,j}^{\dagger} \right),
\label{Eq:DrivenHamiltonianForGeneratingSOC}
\end{eqnarray}
where the driving signals with frequencies $\omega_i$ (with $i=1,2,3,4$) are defined through the subscripts of the coefficients $f$: for example $\omega_1$ corresponds to the signal modulating the coupling strength between the qubit labeled $\downarrow$ and resonator mode 1 at site $i$ (thus corresponding to the bottom left transition in Fig.~\ref{Fig:SpinOrbitCoupling}b).

As shown in Appendix B, with the proper choice of driving-field parameters we obtain the effective Hamiltonian
\begin{eqnarray}
H & = & \tilde{\omega}_{\downarrow} \left( c_{\downarrow,i}^{\dagger} c_{\downarrow,i} + c_{\downarrow,j}^{\dagger} c_{\downarrow,j} \right) + \tilde{\omega}_{\uparrow} \left( c_{\uparrow,i}^{\dagger} c_{\uparrow,i} + c_{\uparrow,j}^{\dagger} c_{\uparrow,j} \right) + \tilde{\omega}_{r,1} \left( a_{1,i}^{\dagger} a_{1,i} + a_{1,j}^{\dagger} a_{1,j} \right) + \tilde{\omega}_{r,2} \left( a_{2,i}^{\dagger} a_{2,i} + a_{2,j}^{\dagger} a_{2,j} \right) \nonumber \\ & & + \tilde{J}_1 \left( a_{1,i}^{\dagger} a_{1,j} + a_{1,i} a_{1,j}^{\dagger} \right) + \tilde{J}_2 \left( a_{2,i}^{\dagger} a_{2,j} + a_{2,i} a_{2,j}^{\dagger} \right) + J_1 \frac{g_{\downarrow,1,i}g_{\downarrow,1,j}}{\Delta_1^2} \left( c_{\downarrow,i}^{\dagger} c_{\downarrow,j} + c_{\downarrow,i} c_{\downarrow,j}^{\dagger} \right) \nonumber \\ & & + J_2 \frac{g_{\uparrow,2,i}g_{\uparrow,2,j}}{\Delta_2^2} \left( c_{\uparrow,i}^{\dagger} c_{\uparrow,j} + c_{\uparrow,i} c_{\uparrow,j}^{\dagger} \right) + J_1 \frac{f_{\downarrow,1,i}f_{\uparrow,1,j}}{4\delta_1^2} \left( e^{-i(\phi_1+\phi_2)} c_{\downarrow,i}^{\dagger} c_{\uparrow,j} + e^{i(\phi_1+\phi_2)} c_{\downarrow,i} c_{\uparrow,j}^{\dagger} \right) \nonumber \\ & & + J_2 \frac{f_{\uparrow,2,i}f_{\downarrow,2,j}}{4\delta_2^2} \left( e^{-i(\phi_3-\phi_4)} c_{\uparrow,i}^{\dagger} c_{\downarrow,j} + e^{i(\phi_3-\phi_4)} c_{\uparrow,i} c_{\downarrow,j}^{\dagger} \right),
\label{Eq:FinalEffectiveHamiltonianSOC}
\end{eqnarray}
\end{widetext}
where the tildes indicate renormalized parameter values, $\Delta_1=\omega_{r,1}-\omega_{\downarrow}$, $\Delta_2=\omega_{r,2}-\omega_{\uparrow}$, $\delta_1=\omega_{r,1}-\omega_1$, and $\delta_2=\omega_{r,2}-\omega_4$. The last four terms in this Hamiltonian along with their tunable parameters allow us to engineer any desired type of effective spin-orbit coupling. In particular, by taking the loop $(\downarrow,i) \; \rightarrow \; (\uparrow,j) \; \rightarrow \; (\uparrow,i) \; \rightarrow \; (\downarrow,j) \; \rightarrow \; (\downarrow,i)$, a particle picks up a phase of $\phi_1+\phi_2+\phi_3-\phi_4$. In the example of the Rashba spin-orbit coupling mentioned above, one can obtain the necessary minus signs by setting the phases to appropriate values, in particular having the combination $\phi_1+\phi_2+\phi_3-\phi_4$ equal to zero for one direction and $\pi$ for the other direction in the two dimensional square lattice. Changing the amplitudes and phases of the driving fields would lead to a continuum of different types of spin-orbit coupling.

State preparation and readout are rather straightforward when
encoding the different spin states in different qubits.
Microwave-controlled quantum operations driven via local antennas
can be used to initialize individual qubits in their excited
states, thus allowing the preparation of well defined numbers of
particles in the different spin states. Similarly the state
readout of individual qubits can be readily achieved in
state-of-the-art setups, and this readout would yield the
positions and spin states of the simulated particles. As discussed in Sec.~\ref{SubSec:PolaritonBranches}, one can also investigate the ground-state phase diagram by preparing theoretically predicted mean-field ground states and monitoring their time evolution.

\section{Experimental parameters}
\label{Sec:ExperimentalParameters}

Superconducting qubits and resonators have typical frequencies in
the range 1-20 GHz. One could therefore think of a resonator that
has a fundamental frequency of about 4 GHz, with a few additional
modes at multiples of this frequency. Qubits with tunable
frequencies in the vicinity of these resonator frequencies can be
fabricated in present-day state-of-the-art experiments.
Qubit-resonator coupling strengths can reach hundreds of
megahertz. One can therefore take 100 MHz as a typical, realistic
value for the coupling strength. Resonator arrays with coupling
strengths of 30 MHz have been fabricated \cite{Underwood}, and
values in the 50-100 MHz range should be possible. Effective
spin-conserving hopping matrix elements on the order of 10 MHz or
higher should therefore be achievable.

For the proposal of Sec.~\ref{SubSec:MultipleQubits}, the single-transition matrix elements in the Raman processes used
in driving the spin-changing hopping processes are limited by the
fact that one needs to keep these transitions virtual. In other
words, one needs to keep the latter's matrix elements smaller than the detuning between the resonator frequency
and the virtual energy level used in the Raman process. This
detuning can be 100 MHz or more. With the single-transition matrix
elements tuned to around 50 MHz, the spin-changing hopping
processes can have effective matrix elements on the order of
10MHz. The fact that both spin-conserving and spin-changing
hopping matrix elements can be engineered in the same range allows
one to design any desired combination of hopping processes and
therefore any desired type of spin-orbit coupling.

Some superconducting qubit designs have weak anharmonicities, and in such cases one normally has to take into consideration the higher energy levels of the qubit circuit. However, the anharmonicity is higher than 200 MHz in most (if not all) experiments to date. Since we focus on the regime where double occupancy is prohibited because of other considerations, and the hopping strength is typically in the range 10-100 MHz (i.e. much smaller than the anharmonicity), the weak anharmonicity of the qubits should not create any additional concerns about the validity of the single-excitation approximation.

Decay rates of superconducting qubits and resonators are steadily
improving (i.e.~decreasing). Decay rates of 10-100 kHz, implying
excitation lifetimes of 10-100 $\mu$s (long compared to the
hopping matrix elements), are now quite realistic. An
implementation of the proposed setup could therefore be realized
in the coming few years.

\section{Conclusion}
\label{Sec:Conclusion}

We have considered the possibility of simulating itinerant
spin-carrying particles using lattices of superconducting qubits
and resonators. The basic JCH setup could be used to realize
a number of relevant processes in this context, while a more complex
system employing multiple qubits coupled to each resonator offers
additional flexibility and could lead to more sophisticated simulations
in the future. In particular, such a system would allow the simulation
of large spin values with possibly improved control of the various
spin-conserving and spin-changing hopping processes.

Experiments on the use of superconducting circuits for implementing
JCH systems are in the early stages of development but are
progressing at a fast pace. They hold promise of great
controllability and measurability, two properties that are highly
desirable in a quantum simulator. We expect that the ability to
add internal degrees of freedom to the simulated particles, along
with the ability to engineer various spin-related physical
processes, will add to the power of this platform for quantum
simulation.

\section*{Acknowledgments}

This work was supported in part by LPS, NSA, ARO, NSF Grant No.
0726909, JSPS-RFBR Contract No. 09-02-92114, Grant-in-Aid for
Scientific Research (S), MEXT Kakenhi on Quantum Cybernetics, and
the JSPS via its FIRST program.

\section*{Appendix A: Two alternative approaches for the simulation of spin-carrying particles}

In this appendix we present two alternative approaches that one might consider for achieving the desired goal of simulating itinerant spin-carrying particles. These approaches might come to mind naturally in the present context. As will be discussed below, however, we find these approaches less promising than the ones presented in the main text.

{\it Higher qubit states} --- Superconducting qubits are in fact
multi-state quantum systems where only two states are used when
representing qubit states. A recent experiment \cite{Neeley} made
use of the additional quantum states in order to simulate a single
spin larger than 1/2. Constructing a resonator array with a single
one of these multi-level circuits coupled to each resonator might therefore seem to be a
promising approach to obtaining the desired system. However, the
energy levels in the qubit circuit are almost (but not exactly) equally
spaced. A similar lack of tunability also arises when considering the
matrix elements that describe the qubit-resonator coupling.
Consequently the tunneling strengths in the resulting JCH model would
be constrained to follow a certain, partially regular pattern, limiting
the ability of the system to probe even the basic parameter regimes of
interest. Constructing the desired quantum simulator using this
system is therefore not as straightforward as it might seem at first
sight.

{\it Multiple resonator modes} --- Owing to their extended structure,
transmission-line resonators (TLRs) generally support a large number of
modes, the so-called fundamental mode with frequency $\omega_f$ and
modes with frequencies that are close to integer-multiples of the
fundamental frequency (i.e.~at frequencies close to $n_m\omega_f$ where
the mode index $n_m=2,3,4,...$). One therefore automatically obtains
multiple (potentially usable) degrees of freedom in a TLR.
Excitations in the different modes can be used to represent particles
in the different internal states. In particular, $2s+1$ modes are
needed in order to simulate spin-$s$ particles. The recently
demonstrated parametric coupling between two modes of a TLR
\cite{ZakkaBajjani} can be used to simulate spin-changing
terms in the Hamiltonian.

There are, however, a number of difficulties associated with
following this approach. The hopping strength in a system with
capacitive coupling between the resonators is proportional to the
capacitive energy between two resonators, which is proportional to
the product of the charges accumulated across the capacitor. For a
given mode $n_m$, each one of the charges is proportional to
$\sqrt{n_m \times n_p}$, where $n_p$ is the number of photons in
mode $n_m$. As a result the hopping strength is proportional to
$n_m$. In order to simulate particles whose hopping strength is
independent of spin, one would like to have no such dependence, or
at least one would like to have tunable coupling strengths with
the ability to set all of them to a single value.

Another complication that arises with the use of multiple TLR
modes is the simple relation between the frequencies of the
different modes. For example, if one drives the system at the
fundamental frequency, the drive signal would be resonant with
multiple processes including single-mode and multi-mode processes
(Note that some of these resonances can be multi-photon resonances
occurring at integer multiples of the driving frequency). One
possible method to circumvent the detrimental effects of this
regular structure of frequencies is to use a combination of modes
that are not integer-multiples of each other, e.g.~use the modes
$n_m=2$ and $n_m=3$ in order to simulate spin-1/2 particles.
However, this solution becomes more demanding for larger spin
values. Because of this and the previously mentioned difficulty,
this approach is also not as straightforward as it might seem at first
sight.

\section*{Appendix B: Derivation of the spin-orbit coupling terms given in Sec.~\ref{SubSec:MultipleQubits}}

In this Appendix we show the derivation of Eq.~(\ref{Eq:FinalEffectiveHamiltonianSOC}) from Eq.~(\ref{Eq:DrivenHamiltonianForGeneratingSOC}). For this derivation we split the problem of obtaining an effective Hamiltonian for the qubit excitations into two separate problems, one where we consider only the effects of the time-independent qubit-resonator coupling terms (with coefficients denoted by the symbol $g$) and one where we consider only the effects of the ac terms (with coefficients denoted by the symbol $f$). By doing so, we assume no interference between these two types of terms in producing the effective Hamiltonian. We shall discuss at the end of the derivation why this approximation is justified.

We first consider the Hamiltonian without the ac terms:
\begin{widetext}
\begin{eqnarray}
H & = & \omega_{\downarrow} \left( c_{\downarrow,i}^{\dagger} c_{\downarrow,i} + c_{\downarrow,j}^{\dagger} c_{\downarrow,j} \right) + \omega_{\uparrow} \left( c_{\uparrow,i}^{\dagger} c_{\uparrow,i} + c_{\uparrow,j}^{\dagger} c_{\uparrow,j} \right) + \omega_{r,1} \left( a_{1,i}^{\dagger} a_{1,i} + a_{1,j}^{\dagger} a_{1,j} \right) + \omega_{r,2} \left( a_{2,i}^{\dagger} a_{2,i} + a_{2,j}^{\dagger} a_{2,j} \right) \nonumber \\ & & + g_{\downarrow,1,i} \left( c_{\downarrow,i}^{\dagger} a_{1,i} + c_{\downarrow,i} a_{1,i}^{\dagger} \right)+ g_{\downarrow,1,j} \left( c_{\downarrow,j}^{\dagger} a_{1,j} + c_{\downarrow,j} a_{1,j}^{\dagger} \right) + g_{\uparrow,2,i} \left( c_{\uparrow,i}^{\dagger} a_{2,i} + c_{\uparrow,i} a_{2,i}^{\dagger} \right) + g_{\uparrow,2,j} \left( c_{\uparrow,j}^{\dagger} a_{2,j} + c_{\uparrow,j} a_{2,j}^{\dagger} \right) \nonumber \\ & & + J_1 \left( a_{1,i}^{\dagger} a_{1,j} + a_{1,i} a_{1,j}^{\dagger} \right) + J_2 \left( a_{2,i}^{\dagger} a_{2,j} + a_{2,i} a_{2,j}^{\dagger} \right).
\end{eqnarray}
We now perform an adiabatic elimination of the resonator modes via the transformation $\tilde{H} = \tilde{U} H \tilde{U}^{\dagger}$, where
\begin{eqnarray}
\tilde{U} & = & \exp \left[ S_1 - S_1^{\dagger} \right] \\
S_1 & = & - \frac{g_{\downarrow,1,i}}{\Delta_1} c_{\downarrow,i}^{\dagger} a_{1,i} - \frac{g_{\downarrow,1,j}}{\Delta_1} c_{\downarrow,j}^{\dagger} a_{1,j} - \frac{g_{\uparrow,2,i}}{\Delta_2} c_{\uparrow,i}^{\dagger} a_{2,i} - \frac{g_{\uparrow,2,j}}{\Delta_2} c_{\uparrow,j}^{\dagger} a_{2,j},
\end{eqnarray}
$\Delta_1=\omega_{r,1}-\omega_{\downarrow}$ and $\Delta_2=\omega_{r,2}-\omega_{\uparrow}$. After truncation to terms that are at least of order $(g/\Delta)^2$ and ignoring non-resonant terms, this transformation results in the effective Hamiltonian
\begin{eqnarray}
\tilde{H} & = & \left( \omega_{\downarrow} - \frac{g_{\downarrow,1,i}^2}{\Delta_1} \right) c_{\downarrow,i}^{\dagger} c_{\downarrow,i} + \left( \omega_{\downarrow} - \frac{g_{\downarrow,1,j}^2}{\Delta_1} \right) c_{\downarrow,j}^{\dagger} c_{\downarrow,j} + \left( \omega_{\uparrow} - \frac{g_{\uparrow,2,i}^2}{\Delta_2} \right) c_{\uparrow,i}^{\dagger} c_{\uparrow,i} + \left( \omega_{\uparrow} - \frac{g_{\uparrow,2,j}^2}{\Delta_2} \right) c_{\uparrow,j}^{\dagger} c_{\uparrow,j} \nonumber \\ & & + \left( \omega_{r,1} + \frac{g_{\downarrow,1,i}^2}{\Delta_1} \right) a_{1,i}^{\dagger} a_{1,i} + \left( \omega_{r,1} + \frac{g_{\downarrow,1,j}^2}{\Delta_1} \right) a_{1,j}^{\dagger} a_{1,j} + \left( \omega_{r,2} + \frac{g_{\uparrow,2,i}^2}{\Delta_2} \right) a_{2,i}^{\dagger} a_{2,i} + \left( \omega_{r2} + \frac{g_{\uparrow,2,j}^2}{\Delta_2} \right) a_{2,j}^{\dagger} a_{2,j} \nonumber \\ & & + J_1 \left[ 1 - \frac{g_{\downarrow,1,i}^2+g_{\downarrow,1,j}^2}{2\Delta_1^2} \right] \left( a_{1,i}^{\dagger} a_{1,j} + a_{1,i} a_{1,j}^{\dagger} \right) + J_2 \left[ 1 - \frac{g_{\uparrow,2,i}^2+g_{\uparrow,2,j}^2}{2\Delta_2^2} \right] \left( a_{2,i}^{\dagger} a_{2,j} + a_{2,i} a_{2,j}^{\dagger} \right) \nonumber \\ & & + J_1 \frac{g_{\downarrow,1,i}g_{\downarrow,1,j}}{\Delta_1^2} \left( c_{\downarrow,i}^{\dagger} c_{\downarrow,j} + c_{\downarrow,i} c_{\downarrow,j}^{\dagger} \right) + J_2 \frac{g_{\uparrow,2,i}g_{\uparrow,2,j}}{\Delta_2^2} \left( c_{\uparrow,i}^{\dagger} c_{\uparrow,j} + c_{\uparrow,i} c_{\uparrow,j}^{\dagger} \right).
\label{Eq:EffectiveHamiltonianSpinConservingTerms}
\end{eqnarray}
All but the last two terms in the above Hamiltonian correspond to terms in the original Hamiltonian with some small shifts in the effective parameters from the original values (Note that the shifts can be set to equal values, such that they do not detune previously resonant energy levels). The last two terms in the new Hamiltonian describe processes that were not explicitly present in the original Hamiltonian, namely the hopping of $\uparrow$ and $\downarrow$ excitations between neighbouring sites. Note that these processes conserve the spin of the hopping particles. It is also important to emphasize here that the appearance of these terms was the result of applying a transformation that eliminated (to lowest order) the qubit-resonator coupling terms from the Hamiltonian.

Next we consider the Hamiltonian with the ac terms (and without the time-independent coupling terms, whose effect has already been calculated). We shall set $\omega_1+\omega_2=\omega_4-\omega_3=\omega_{\uparrow}-\omega_{\downarrow}$ in order to drive the desired spin-changing processes. We also set $\omega_{\downarrow}=0$ as an energy reference in order to simplify the expressions below. We now perform a rotating-frame transformation $H' = U H U^{\dagger}$, where
\begin{equation}
U = \exp \left\{ -i \left[ \omega_1 \left( a_{1,i}^{\dagger} a_{1,i} + a_{1,j}^{\dagger} a_{1,j} \right) + \omega_{\uparrow} \left( c_{\uparrow,i}^{\dagger} c_{\uparrow,i} + c_{\uparrow,j}^{\dagger} c_{\uparrow,j} \right) + \omega_4 \left( a_{2,i}^{\dagger} a_{2,i} + a_{2,j}^{\dagger} a_{2,j} \right) \right] t\right\}.
\end{equation}
After applying the rotating-wave approximation, i.e.~ignoring terms that oscillate with frequencies that are on the order of the driving ac field frequencies, we obtain the Hamiltonian
\begin{eqnarray}
H' & = & \delta_1 \left( a_{1,i}^{\dagger} a_{1,i} + a_{1,j}^{\dagger} a_{1,j} \right) + \delta_2 \left( a_{2,i}^{\dagger} a_{2,i} + a_{2,j}^{\dagger} a_{2,j} \right) + \frac{f_{\downarrow,1,i}}{2} \left( e^{-i \phi_1} c_{\downarrow,i}^{\dagger} a_{1,i} + e^{i \phi_1} c_{\downarrow,i} a_{1,i}^{\dagger} \right) \nonumber \\ & & + \frac{f_{\uparrow,1,j}}{2} \left( e^{i \phi_2} c_{\uparrow,j}^{\dagger} a_{1,j} + e^{-i \phi_2} c_{\uparrow,j} a_{1,j}^{\dagger} \right) + \frac{f_{\uparrow,2,i}}{2} \left( e^{-i \phi_3} c_{\uparrow,i}^{\dagger} a_{2,i} + e^{i \phi_3} c_{\uparrow,i} a_{2,i}^{\dagger} \right) \nonumber \\ & & + \frac{f_{\downarrow,2,j}}{2} \left( e^{-i \phi_4} c_{\downarrow,j}^{\dagger} a_{2,j} + e^{i \phi_4} c_{\downarrow,j} a_{2,j}^{\dagger} \right) + J_1 \left( a_{1,i}^{\dagger} a_{1,j} + a_{1,i} a_{1,j}^{\dagger} \right) + J_2 \left( a_{2,i}^{\dagger} a_{2,j} + a_{2,i} a_{2,j}^{\dagger} \right),
\end{eqnarray}
where $\delta_1=\omega_{r,1}-\omega_1$, $\delta_2=\omega_{r,2}-\omega_4$. We now perform an adiabatic elimination of the resonator modes via the transformation $\tilde{H}' = \tilde{U}' H' \tilde{U}'^{\dagger}$, where
\begin{eqnarray}
\tilde{U}' & = & \exp \left[ S_2 - S_2^{\dagger} \right] \\
S_2 & = & - \frac{f_{\downarrow,1,i}}{2\delta_1} e^{-i \phi_1} c_{\downarrow,i}^{\dagger} a_{1,i} - \frac{f_{\uparrow,1,j}}{2\delta_1} e^{i \phi_2} c_{\uparrow,j}^{\dagger} a_{1,j} - \frac{f_{\uparrow,2,i}}{2\delta_2} e^{-i \phi_3} c_{\uparrow,i}^{\dagger} a_{2,i} - \frac{f_{\downarrow,2,j}}{2\delta_2} e^{-i \phi_4} c_{\downarrow,j}^{\dagger} a_{2,j}.
\end{eqnarray}
After truncation to terms that are at least of order $(f/\delta)^2$ and ignoring non-resonant terms, we obtain the effective Hamiltonian
\begin{eqnarray}
\tilde{H}' & = & \left( \delta_1+\frac{f_{\downarrow,1,i}^2}{4\delta_1^2} \right) a_{1,i}^{\dagger} a_{1,i} + \left( \delta_1+\frac{f_{\uparrow,1,j}^2}{4\delta_1^2} \right) a_{1,j}^{\dagger} a_{1,j} + \left( \delta_2+\frac{f_{\uparrow,2,i}^2}{4\delta_2^2} \right) a_{2,i}^{\dagger} a_{2,i} + \left( \delta_2+\frac{f_{\downarrow,2,j}^2}{4\delta_2^2} \right) a_{2,j}^{\dagger} a_{2,j} \nonumber \\ & & - \frac{f_{\downarrow,1,i}^2}{4\delta_1^2} c_{\downarrow,i}^{\dagger} c_{\downarrow,i} - \frac{f_{\downarrow,2,j}^2}{4\delta_2^2} c_{\downarrow,j}^{\dagger} c_{\downarrow,j} - \frac{f_{\uparrow,2,i}^2}{4\delta_2^2} c_{\uparrow,i}^{\dagger} c_{\uparrow,i} - \frac{f_{\uparrow,1,j}^2}{4\delta_1^2} c_{\uparrow,j}^{\dagger} c_{\uparrow,j} \nonumber \\ & & + J_1 \left[ 1 - \frac{f_{\downarrow,1,i}^2+f_{\uparrow,1,j}^2}{8\delta_1^2} \right] \left( a_{1,i}^{\dagger} a_{1,j} + a_{1,i} a_{1,j}^{\dagger} \right) + J_2 \left[ 1 - \frac{f_{\uparrow,2,i}^2+f_{\downarrow,2,j}^2}{8\delta_2^2} \right] \left( a_{2,i}^{\dagger} a_{2,j} + a_{2,i} a_{2,j}^{\dagger} \right) \nonumber \\ & & + J_1 \frac{f_{\downarrow,1,i}f_{\uparrow,1,j}}{4\delta_1^2} \left( e^{-i(\phi_1+\phi_2)} c_{\downarrow,i}^{\dagger} c_{\uparrow,j} + e^{i(\phi_1+\phi_2)} c_{\downarrow,i} c_{\uparrow,j}^{\dagger} \right) + J_2 \frac{f_{\uparrow,2,i}f_{\downarrow,2,j}}{4\delta_2^2} \left( e^{-i(\phi_3-\phi_4)} c_{\uparrow,i}^{\dagger} c_{\downarrow,j} + e^{i(\phi_3-\phi_4)} c_{\uparrow,i} c_{\downarrow,j}^{\dagger} \right).
\label{Eq:EffectiveHamiltonianSpinChangingTerms}
\end{eqnarray}
\end{widetext}
Once again, all but the last two terms describe small renormalization effects to the Hamiltonian parameters. The last two terms describe spin-changing hopping processes. Combining the results of Eqs.~(\ref{Eq:EffectiveHamiltonianSpinConservingTerms}) and (\ref{Eq:EffectiveHamiltonianSpinChangingTerms}), we obtain Eq.~(\ref{Eq:FinalEffectiveHamiltonianSOC}).

We now go back to the question of whether splitting the problems into smaller (and separate) parts is justified. In particular, the eight different transitions contributing to the four hopping processes could in principle produce additional combinations that did not appear when we divided the driving fields into two separate groups. However, if the inter-site hopping strengths $J$ are much smaller than the detunings $\Delta$ and $\delta$, all of these additional combinations will describe non-resonant Raman processes that can be ignored. One must also be careful about such interference effects when generalizing the above construction to an array of lattice sites. For example, if the same frequency combinations are used to drive the transitions between sites $i$ and $i+1$ and between sites $i+1$ and $i+2$, then undesirable transitions will be driven as well. This problem can be avoided by using different frequencies (or in other words different virtual energy levels) for the different pairs of neighbouring lattice sites.


\begin{thebibliography}{99}
\bibitem{HalfIntegerBosonFootnote} It should be noted here that in the JCH
setup we naturally obtain effective particles with bosonic statistics.
Although truly spin-1/2 particles are fermions, one can think of
pseudospin-1/2 bosonic particles. Such particles have for example been
considered extensively in cold atomic gases \cite{SpinorBEC}.
\bibitem{DispersiveRegimeFootnote} In addition to the dispersive regime, another commonly studied special case in the the JCH setup is the that where the qubits and resonators have the same frequency. In this case the energy levels exhibit a different dependence on the number of excitation quanta. However, if one assumes a dilute system, where it suffices to consider at most double occupancy of a given site, any on-site energy level structure can be described by the interaction term given in Eq.~(\ref{Eq:BH_Hamiltonian_WithoutSpin}).
\bibitem{DrivingFieldPhaseFootnote} Note that the field direction is also
determined by the initial-time phase of the driving fields. This phase might
have no physical consequences in the single-site case, but the relation between
the phases at different sites in the lattice determines the profile of the
simulated external field.
\bibitem{HatsAndHbarFootnote} In this derivation we shall not explicitly include the hat symbol for the operators, and we set $\hbar=1$.
\end{thebibliography}
\end{document}